\documentclass[aps,showpacs,eqsecnum,floatfix]{revtex4}

\usepackage[dvips]{graphicx,color}

\begin{document}
\title{Faddeev and Glauber Calculations at Intermediate
Energies in a Model for n+d
Scattering}

\author{Ch.~Elster$^{(a)}$}

\author{T.~Lin$^{(a)}$}

\author{W.~Gl\"ockle$^{(b)}$}

\author{S.~Jeschonnek$^{(c)}$}

\affiliation{(a)
Institute of Nuclear and Particle Physics,  and
Department of Physics and Astronomy,  Ohio University,
Athens, OH 45701,
USA}
\affiliation{(b)
Institute for Theoretical Physics II, Ruhr-University Bochum,
D-44780 Bochum, Germany}
\affiliation{(c)
Department of Physics, The Ohio State University,
Lima, OH 45804, USA}

\vspace{10mm}

\date{\today}
\vspace{5mm}

\begin{abstract}
Obtaining cross sections for nuclear reactions at intermediate
energies based on the Glauber formulation has a long tradition. Only
recently the energy regime of a few hundred MeV has become
accessible to ab-initio Faddeev calculations of three-body
scattering. In order to go to higher energies, the Faddeev
equation for three-body scattering is formulated and directly solved
without employing a partial wave decomposition.
In the simplest form the Faddeev equation for interacting
scalar particles is a three-dimensional integral equation in five
variables, from which the total cross section, the cross sections
for elastic scattering and breakup reactions,
as well as differential cross sections are obtained.
The same observables are calculated based on the Glauber formulation.
The first order
Glauber calculation and the Glauber rescattering corrections
are compared  in detail
with the corresponding terms of the Faddeev multiple scattering
series for projectile energies between 100 MeV and 2 GeV.
\end{abstract}
\vspace{10mm}

\pacs{21.45-v,24.10.Ht,25.10.+s}

\maketitle


\section{Introduction}

The formulation of high energy scattering introduced by Glauber
\cite{Glauber1,Glauber2} has a long
tradition in being applied to calculating nuclear reactions with
hadronic as well as electromagnetic probes.  Today Glauber amplitudes
are widely used to account for final state interactions in (e,e'p)
reactions at high and intermediate
energies~\cite{Ciofi_1,schiavilla,bianconi}. Glauber theory is applied
to investigate color transparency, in $(e,e'p)$ reactions and in the
electroproduction of mesons \cite{frankfurt,nikolaev}. In addition,
it is applied to
heavy ion reactions \cite{heavyion}. Most recently, several
generalizations of Glauber theory have been employed for the
description of electron scattering mainly from lighter nuclei,
e.g. the generalized eikonal approximation (GEA)
\cite{sargsian,ciofi2}, or relativistic eikonal approaches
\cite{gent_relat_eikonal,ryckeb}. These methods are applied both to
$A(e,e'p)$ and $A(p,2p)$ reactions.
Nuclear scattering at intermediate energies
in the context of radioactive beams also takes advantage
of this formulation~\cite{Tostevin1,Tostevin2}.

Despite the widespread use of the Glauber formulation in reactions involving
few as well as many-body systems there has been little work on rigorous
tests of the accuracy and/or limits of the Glauber ansatz, apart from some
studies concerning the quality of the eikonal approximation for $(e,e'p)$
reactions~\cite{bianconi2}  as well as  for elastic proton
scattering from halo nuclei~\cite{Crespo:2007zza}.

Only recently exact Faddeev calculations for three-body scattering
in the intermediate energy regime
became available. This progress is based on a formulation
of the Faddeev equations, which is based directly on momentum variables
and does not rely on traditional partial wave expansions. The formulation
and numerical realization for the nonrelativistic Faddeev
equations~\cite{Liu:2004tv,Liu:2002gh}
as well as fully Poincar{\'e} invariant
ones~\cite{Lin:2007ck,Lin:2007kg,Lin:2008sy} have been carried out for scalar
interactions up to projectile energies of 2~GeV.
This now allows us to perform a detailed comparison of
three-body Faddeev calculations
of total and elastic cross sections with the ones obtained by
Glauber calculations, and thus test the range of validity of Glauber
calculations at least in the three-body system.

The naive expectation is that for sufficiently high energies
the total $n+d$ cross section is just
the sum of the two total cross sections for two-nucleon scattering.
The question whether this is valid  is an old one
and has been investigated in the context of the eikonal
approximation~\cite{Glauber1,Harr64,FrancoGlauber,BJM}.
It turns out that rescattering (shadowing) corrections need to be
added.
Since at the time no exact calculations were available,
the question remained open,
whether those corrections were sufficient and at which energy the eikonal
approximation became valid.
We will investigate such questions
in this work and mostly follow the formulation given in
Refs.~\cite{Harr64,FrancoGlauber}. In order to make closer contact
with that work as well as for transparency,
we will use the nonrelativistic formulation of the Faddeev
equation for our comparison, even though we carry out calculations in the
intermediate energy regime.  We will also use the same two-body scattering
amplitude as well as deuteron wave function
in the Glauber calculation as enters in our Faddeev results, so
that we can clearly identify the effects of the Glauber approximation
on the multiple scattering.
In Section~II we will briefly present the Faddeev framework and
its multiple scattering
expansion, and in Section~III we will re-derive the essential
expressions of the Glauber
approximation necessary for the comparison to the Faddeev formulation.
Our numerical results for the total and differential cross section for elastic
scattering in both formulations and their discussion will be given in Section~IV.
We summarize and conclude in Section~V.

\section{The Faddeev Multiple Scattering Series}
Various presentations of three-body scattering in the Faddeev scheme
are presented in the literature~\cite{wgbook,wgphysrep}.
We solve the Faddeev equation for three identical particles in the
form
\begin{equation}
T|\phi\rangle=tP|\phi\rangle+tPG_{0}T|\phi\rangle. \label{eq:2.1}
\end{equation}
The driving term of this integral equation consists of the two-body
t-matrix $t$, the sum $P=P_{12}P_{23}+P_{13}P_{23}$ of a cyclic and
anti-cyclic permutation of three particles, and the
initial state $|\phi\rangle = |\phi_d {\bf q_0}\rangle$, composed of
a two-body bound state $\phi_d$
and the momentum eigenstate of the projectile particle.
The kernel of Eq.~(\ref{eq:2.1}) contains the free three-body
propagator,
$G_{0}=(E-H_{0}+i\varepsilon)^{-1}$,
where $E$ is the total energy in the center-of-momentum (c.m.) frame.

\noindent
The operator $T$ determines both, the full break-up amplitude
\begin{equation}
U_{0}=(1+P)T ,
\label{eq-U0}
\end{equation}
and the amplitude for elastic scattering
\begin{equation}
U=PG^{-1}_{0}+PT.
\label{eq-U}
\end{equation}
In this paper we focus on three identical bosons and use a momentum
space representation.
For solving Eq.~(\ref{eq:2.1}), we introduce the standard Jacobi momenta
{\bf p} for the relative momentum
in the subsystem, and {\bf q} for the relative momentum of the spectator
to the subsystem.
The momentum states are normalized according to
$\langle {\bf p}'{\bf q}'|{\bf p} {\bf q}\rangle =
\delta^3 ({\bf p}'-{\bf p}) \delta^3 ({\bf q}' -{\bf q})$.
Projecting  Eq.~(\ref{eq:2.1}) onto Jacobi momenta leads to
\cite{Liu:2004tv}
\begin{eqnarray}
\langle{\mathbf{p}}{\mathbf{q}}|T|\phi_{d}{\mathbf{q}}_{0}\rangle
&=&
\phi_{d}\left({\mathbf{q}}+\frac{1}{2}{\mathbf{q}}_{0}\right)
t_{s}\left({\mathbf{p}},\frac{1}{2}{\mathbf{q}}+{\mathbf{q}}_{0},E-
\frac{3}{4m}q^{2} \right) \nonumber \\
&+&\int
d^{3}q''t_{s}\left({\mathbf{p}},\frac{1}{2}{\mathbf{q}}+{\mathbf{q}}'',
E-\frac{3}{4m}q^{2}\right)
\frac
{\left\langle{\mathbf{q}}+\frac{1}{2}{\mathbf{q}}'',{\mathbf{q}}''
\left|T\right| \phi_{d}{\mathbf{q}}_{0} \right\rangle}
{E-\frac{1}{m}(q^{2}+q''^{2}+{\mathbf{q}}\cdot{\mathbf{q}}'')+i\varepsilon}.
\label{eq:2.4}
\end{eqnarray}
Here
$t_{s}({\mathbf{p}}',{\mathbf{p}})=
t({\mathbf{p}},{\mathbf{p}}')+t(-{\mathbf{p}}',{\mathbf{p}})$
is the symmetrized two-body $t$ matrix and the total energy $E$ is
explicitly given as
\begin{equation}
 E=E_{d}+\frac{3}{4m}q^{2}_{0} =E_{d}+\frac{2}{3}E_{lab}.
\label{eq:2.5}
\end{equation}
We assume that the underlying two-body force  generates a $t$-matrix
as solution of a two-body Lippmann-Schwinger equation, and
that the force supports one bound state with energy $E_d$.
Then Eq.~(\ref{eq:2.4})  is solved directly in
momentum space as function of the vector Jacobi momenta and angles
between them. For further details we refer to Ref.~\cite{Liu:2004tv}.

In this work we want to study the high energy behavior of the
cross sections. In the framework of the Faddeev formulation, we
consider the multiple scattering series defined by the integral
equation given in Eq.~(\ref{eq:2.1}).
For the elastic scattering amplitude this leads to the series
expansion
\begin{equation}
\langle\Phi| U | \Phi\rangle = \langle \Phi|PG_0^{-1}| \Phi\rangle +
\langle \Phi|PtP| \Phi\rangle + \langle \Phi|PtPG_0tP| \Phi\rangle
+\cdots ,
\label{eq:2.6}
\end{equation}
which for the fully converged solution is summed as Pad\'e or Neumann
series.
 Via the optical
theorem the total cross section is related to the imaginary part of the
operator $U$ as
\begin{equation}
\sigma_{tot}^{\rm ND}=-(2\pi)^3 \frac{4m}{3 q_0} \Im m \langle \Phi|U|\Phi
\rangle.
\label{eq:2.7}
\end{equation}
This again can be expanded in a multiple scattering series through
\begin{eqnarray}
2 i \; \Im m \langle \Phi|U|\Phi\rangle &=& \langle \Phi|U|\Phi\rangle
-\langle \Phi|U|\Phi\rangle^\star  \nonumber \\
&=& \langle \Phi|P(t-t^\dagger) P|
\Phi\rangle  +\langle \Phi|PtPG_0tP| \Phi\rangle
 -\langle \Phi|P t^\dagger G_0^* Pt^\dagger P| \Phi\rangle + \cdots
\label{eq:2.8}
\end{eqnarray}
It turns out that the Faddeev multiple scattering series for the total
cross sections converges fast for projectile energies higher than
500~MeV \cite{Liu:2004tv,Lin:2008sy}.
This finding is encouraging for our study of the Glauber formulation.

One can also start from  Eq.~(\ref{eq:2.8}) and  derive a
high-energy limit in the frame work of the Faddeev equations.
This has been carried out in Ref.~\cite{Elster:1998ry}.
Here the terms of first and second order in $t$ have  been investigated
in the limit $ E\rightarrow\infty$ leading to the analytic result
\begin{eqnarray}
\sigma_{tot}^{\rm ND} &=& 2 \sigma_{tot}^{\rm NN} \nonumber \\
&+& (2\pi)^5 (\frac{4m }{3q_0})^2 \left[\Re e \langle \frac{3}{4}
{\bf q_0} | \; t(\frac{1}{m}( \frac{3}{4} q_0)^2 ) \; | \frac{3}{4}
{\bf q_0}\rangle \right]^2
\langle \phi_d|\frac{1}{r^2} | \phi_d\rangle \frac{1}{4 \pi}
-( \sigma_{tot}^{\rm NN})^2
\langle \phi_d| \frac{1}{r^2} | \phi_d\rangle \frac{1}{4 \pi}
\nonumber \\
&=& 2 \sigma_{tot}^{\rm NN} + O(t^2) + O(t^4)
\label{eq:2.9}
\end{eqnarray}
In first order this expansion gives twice the two-nucleon cross
section. The second order correction terms can now be tested against
exact Faddeev calculations. In Ref.~\cite{Elster:1998ry} this was not
possible.

\section{The Glauber Calculation}
For the convenience of the reader we briefly sketch the derivation of
the Glauber amplitude for
elastic $n+d$ scattering, following the work given in
Refs.~\cite{Harr64,FrancoGlauber}.
In the laboratory system with the projectile momentum $\bf k$
and the momentum transfer $\bf q  = \bf k - {\bf k}'$ the two- particle
scattering amplitude according to Glauber~\cite{Glauber1} is assumed to have the
form
\begin{eqnarray}
f_k ({\bf q}) = \frac{ik}{2 \pi} \int d^2 b \; e^{i {\bf q}\cdot{\bf b}}
 \;\Gamma({\bf b})\label{eq3.1}
 \end{eqnarray}
where $ \Gamma ({\bf b}) $ is given through the eikonal phase
$\chi({\bf b})$ as
\begin{eqnarray}
\Gamma ({\bf b})  = 1 - e^{i \chi({\bf b})}.
\end{eqnarray}
The vector ${\bf b}$ is assumed to be  perpendicular to a direction $ \hat n$,
specified below. Scattering off a deuteron target then leads to
\begin{eqnarray}
F_k({\bf q})  =   \frac{ik}{2 \pi} \int d^2 b \; e^{i{\bf q} \cdot
{\bf b}} \int d^3 r \; \phi_d^* ({\bf r})
 ( 1 - e^{i( \chi({\bf b}-1/2{\bf s}) + \chi ({\bf b} + 1/2{\bf s})) })
 \; \phi_d({\bf r}),
\end{eqnarray}
where $\phi_d({\bf r}) $ is the deuteron wave function and ${\bf r} =
{\bf s} + \hat n ( \hat n \cdot{\bf r}) $. Using these definitions it follows that
\begin{eqnarray}
F_k({\bf q}) & = &  \frac{ik}{2 \pi} \int d^2 b \; e^{i{\bf q} \cdot
{\bf b}} \int d^3 r \phi_d^* ({\bf r})\nonumber\\
& & \left[
\Gamma ({\bf b} - \frac{1}{2}{\bf s}) + \Gamma({\bf b} + \frac{1}{2}{\bf s}) -
  \Gamma({\bf b}-\frac{1}{2}{\bf s}) \Gamma({\bf b} +\frac{1}{2}{\bf s})\right]
\phi_d({\bf r})
\end{eqnarray}
Using the inverse of Eq.~(\ref{eq3.1}),
\begin{equation}
 \Gamma({\bf b}) = \frac{1}{2 \pi i k } \int d^2 q \; e^{- i{\bf q} \cdot
{\bf b}}f_k({\bf q}),
\label{Gam}
 \end{equation}
and again integrating over a vector ${\bf q}$ in a plane
 perpendicular to $\hat n$ one
can eliminate the $ \Gamma$'s in favor of the two-particle scattering
amplitude leading to the expression given in Refs.~\cite{Harr64,FrancoGlauber}
\begin{eqnarray}
F_k({\bf q})    =   2 S( \frac{1}{2}{\bf q}) f({\bf q})
 + \frac{i }{2 \pi k} \int d^2 q' S({\bf q'})
f_k( \frac{1}{2}{\bf q} +{\bf q'}) f_k ( \frac{1}{2}{\bf q} -{\bf q'}).
\label{eq3.6}
\end{eqnarray}
The deuteron wave function occurs in the  form factor
\begin{eqnarray}
S({\bf q}) = \int d^3 r \vert \phi_d({\bf r})\vert ^2 e^{i{\bf q} \cdot{\bf s} }
 =\int d^3 r \vert \phi_d({\bf r})\vert ^2 e^{i{\bf q} \cdot{\bf r} }.
\label{deutformf}
\end{eqnarray}
From the above expression one obtains the laboratory
differential cross section for $n+d$ scattering as
\begin{eqnarray}
\frac{d \sigma^{{\rm ND}}}{d \Omega_l} = | F_k({\bf q})|^2
\label{eq3.8}
\end{eqnarray}
and the total cross section for $n+d$ scattering
\begin{eqnarray}
\sigma^{{\rm ND}}_{tot} = \frac{4 \pi}{k} \Im m F_k ({\bf 0})
\label{eq3.9}
\end{eqnarray}

\noindent
In first order in the two-body scattering amplitude Eq.~(\ref{eq3.9})
leads to the expected result for the total cross section
\begin{eqnarray}
\sigma^{{\rm ND},1st}_{tot}= 2 S(0) \frac{4 \pi}{k}  \Im m f_k(0) =
  2 \frac{4 \pi}{k}   \Im m f_k(0) =  2 \sigma^{{\rm NN}}_{tot}
\label{eq3.10}
\end{eqnarray}
and to the differential cross section for elastic  $n+d$ scattering
\begin{eqnarray}
\frac{d \sigma^{{\rm ND},1st}}{d \Omega_l} = 4 S^2 \left( \frac{1}{2}{\bf q}\right)
\vert f_k({\bf q})\vert ^2
\label{eq3.11}
\end{eqnarray}

For the total cross section for $n+d$ scattering, including the second order
correction, one obtains
\begin{eqnarray}
\sigma^{{\rm ND}}_{tot} = 2 \sigma^{{\rm NN}}_{tot} + \delta \sigma_{tot},
\label{eq3.12}
\end{eqnarray}
where
\begin{eqnarray}
\delta \sigma_{tot} = \frac{2}{k^2} \int d^2 q' \; S({\bf q'})
 f_k({\bf q'} ) f_k( -{\bf q'}).
\label{eq3.13}
\end{eqnarray}
This correction term represents rescattering events.

\noindent
Adding the rescattering corrections to the differential cross section for
elastic scattering leads to
\begin{eqnarray}
 \frac{d \sigma^{{\rm ND}}}{d \Omega_l} &=& 4 S^2 \left( \frac{1}{2}{\bf q}\right)
\; \vert f_k({\bf q})\vert^2\cr
&-& \frac{2}{\pi k} S\left(\frac{1}{2}{\bf q}\right) \; \Im m \; \left[ f_k^*({\bf q})
 \int d^2 q' \; S({\bf q'}) \; f_k( 1/2{\bf q}+{\bf q'}) \;
f_k( 1/2{\bf q} -{\bf q'}) \right]\cr
&+ &  \left( \frac{1}{2 \pi k}\right)^2 \left| \int d^2 q' \; S({\bf q'})
 \; f_k( 1/2{\bf q} +{\bf q'}) \; f_k( 1/2{\bf q} - {\bf q'})\right|^2.
\label{eq3.14}
\end{eqnarray}

Next, we need to discuss the choice of the two-particle scattering amplitude
$f_k({\bf q})$. In the original work of Refs.~\cite{Glauber1,Harr64,FrancoGlauber},
the eikonal phase is given in terms of the
 potential $V({\bf r})$ as
\begin{eqnarray}
\chi({\bf b})& = &  - \frac{1}{v}\int_{-\infty}^{\infty} V({\bf b} +{\bf z}) \; dz\cr
&  = &  - \frac{2m}{k}\int_0^{\infty} V(\sqrt{b^2 + z^2}) \; dz,
\end{eqnarray}
Where ${\bf z}$ is a vector parallel to ${\bf k}$. The above definition would result
in a two-body scattering amplitude in the eikonal approximation. Studies of
the quality of the eikonal approximation for potential scattering
have been carried out in detail in the
past for different potentials. The for us interesting case of Yukawa-type
potentials is considered e.g. in Refs.~\cite{BJM,Wallacean}. Since our main interest
is the comparison of the $n+d$ cross sections, we need to use the same two-particle
scattering amplitude $f_k({\bf q})$ for our Glauber calculation as we use for the
Faddeev calculation. Thus we need to use the scattering amplitude obtained from
the solution of a Lippmann-Schwinger equation with the potential $V$ as driving term.

This two-particle scattering amplitude
$f_c({\bf p'},{\bf p})$ is given in the c.m. frame
and is related to the on-shell two-body $t$-matrix as
\begin{equation}
\langle{\bf p'} | V | \psi_{{\bf p}}^{(+)}\rangle =
\langle {\bf p'} |t (E_p=\frac{p^2}{m}|{\bf p} \rangle  \equiv
t({\bf p'},{\bf p}; \frac{p^2}{m}) ,
\end{equation}
and leads to the scattering amplitude
\begin{eqnarray}
f_c({\bf p'},{\bf p}) =-\frac{m}{2} (2\pi)^2 \; t({\bf p'},{\bf p};\frac{p^2}{m}).
\end{eqnarray}
For a scalar potential $t$ is also a scalar and depends on the magnitude of
$|{\bf p}|=|{\bf p'}|$ and the angle $x_c= {\hat {\bf p'}}\cdot {\hat {\bf p}}$ between
the two vectors ${\bf p'}$ and ${\bf p}$ \cite{3dt},
\begin{eqnarray}
t({\bf p'},{\bf p}; \frac{p^2}{m}) = t( p, p, x_c; \frac{p^2}{m}) \equiv t( p,x_c).
\label{eq3.18}
\end{eqnarray}
For identical particles we use as in the Faddeev calculations the
symmetrized $t$-matrix $t_s({\bf p'},{\bf p};\frac{p^2}{m})$.
The c.m. differential cross section for the scattering of two identical
particles in then given as
\begin{eqnarray}
\frac{d \sigma}{d \Omega_c} = | f^s_c( p,x_c)|^2
\end{eqnarray}
and the laboratory differential cross section as
\begin{eqnarray}
\frac{d \sigma}{d \Omega_l} = 4 x_l \frac{d \sigma}{d \Omega_c},
\end{eqnarray}
 where the standard relation between c.m. and laboratory scattering
 angles for two equal mass particles is given as
\begin{eqnarray}
x_l \equiv {\hat {\bf  k}} \cdot {\hat {\bf k'}} &=&
\sqrt{\frac{x_c + 1}{2}} \nonumber \\
x_c &=&  2 x_l^2 - 1.
\label{eq3.20}
\end{eqnarray}
If one defines
\begin{eqnarray}
\frac{d \sigma}{d \Omega_l} = | f^s_l( k,x_l)|^2
\end{eqnarray}
 the two-particle scattering amplitude in the laboratory frame reads
\begin{eqnarray}
f^s_l( k,x_l) = 2 \sqrt{x_l} f^s_c( p,x_c(x_l)).
\label{eq3.23}
\end{eqnarray}
All that remains now is  choosing the unit vector $\hat n$.
The original suggestion by Glauber \cite{Glauber1,FrancoGlauber} is
\begin{eqnarray}
{\rm (a)}: \;\;  {\hat n} = {\hat {\bf  k}},
\label{eq3.24}
\end{eqnarray}
with the approximation $ |{\bf k}|  = |{\bf k'}|$.
In other works~\cite{Harr64,BJM,Wallacean} the choice
\begin{eqnarray}
{\rm (b)}: \;\; \hat n = \widehat{{\bf k}+{\bf k'}}
\label{eq3.25}
\end{eqnarray}
is preferred.
We will use both choices in our calculations to explore the differences.
The momentum transfer ${\bf q}$ enters the scattering amplitude $f_c$ in the form of
\begin{eqnarray}
x_c = 1 - \frac{{\bf q}^2}{ 2 p^2}
\label{eq3.26}
\end{eqnarray}
The  Glauber ansatz requires to replace the momentum
transfer ${\bf q}^2$ by  $ {\bf q_{\perp}}^2$
with $ \bf q_{\perp} $ being orthogonal to $\hat n$. For the two cases of
Eqs.~(\ref{eq3.24}) and (\ref{eq3.25}) this leads to
\begin{eqnarray}
{\rm (a)} : \;\; {\bf q_{\perp}}^2 & = &  k^2 ( 1- x_l^2) \nonumber \\
{\rm (b)} : \;\; {\bf q_{\perp}}^2  & = &  k^2 (1-x_l^2) \frac{4 x_l^2}{1+ 3x_l^2}.
\label{eq3.28}
\end{eqnarray}
For forward angles both choices become equivalent.

\noindent
The total cross section for two-body scattering entering
into Eq.~(\ref{eq3.10}) is then explicitly given as
\begin{eqnarray}
\sigma^{{\rm NN}}_{tot} = \int d\Omega_c | f^s_c( p,x_c)|^2.
\end{eqnarray}
The scattering amplitude $f_k({\bf q})$ entering Eq.~(\ref{eq3.11})
can now be directly obtained  from  Eq.~(\ref{eq3.23}) as
\begin{eqnarray}
f_k({\bf q}) \rightarrow f^s_l( k,x_l).
\end{eqnarray}
In calculating the angle $x_c(x_l)$, the momentum transfer of Eq.~(\ref{eq3.26}) must
then be replaced by $ q_{\perp}^2$,
and $ x_l$ is related to $x_c$ through Eq.~(\ref{eq3.20}).

For the second order correction to the differential cross section the
integral in Eq.~(\ref{eq3.14}) needs to be evaluated. According to the
Glauber ansatz the argument of the scattering amplitude needs to be
evaluated as $1/2 {\bf q}_\perp \pm {\bf q'}$, so that the integration is carried out
in the plane defined by ${\hat n}$.
Explicitly, the integral is calculated as
\begin{eqnarray}
\int d^2 q' & & S(q') f^s_k(\frac{1}{2}{\bf q} +{\bf q'})
f^s_k(\frac{1}{2}{\bf q} -{\bf q'}) \rightarrow \cr
& &4 \left(\frac{m}{2}\right)^2 (2 \pi)^4
 \int_0^{q_{max}} d q' q' \; S(q') \int_0^{2 \pi} d\phi
\;  \sqrt{x_l(x^{(+)})x_l(x^{(-)})} \; t_s( p,x^{(+)})  t_s( p,x^{(-)}),
\label{eq3.32}
\end{eqnarray}
with the definition of $t_s(p,x)$ from Eq.~(\ref{eq3.18}) and
\begin{eqnarray}
x^{(\pm)} = 1- \frac{1/4 q_{\perp}^2 \pm q_{\perp} q' \cos \phi + {q'}^2}{2 p^2}.
\end{eqnarray}
Here  $\phi$ is the angle between the vectors ${\bf q}_{\perp}$ and $\bf q'$.
The upper limit
$q_{max}$ is the maximum momentum transfer allowed for a given projectile laboratory
energy, corresponding to $x_c=-1$ in the angle argument of the $t$-matrix.

\noindent
For the second order correction to the total cross section, Eq.~(\ref{eq3.12}),
we only need to consider the special case $q_{\perp}=0$.
Then the integral of Eq.~(\ref{eq3.32}) simplifies to
\begin{eqnarray}
\delta \sigma_{tot}  = \frac{2}{k^2} m^2 (2 \pi)^5
\int_0^{q_{max}} d q' \; q' S(q~') x_l(x) \; \Re e \left[t_s^2( p,x) \right],
\label{eq3.34}
\end{eqnarray}
with $x=1-\frac{q'^2}{2p^2}$.

In Ref.~\cite{FrancoGlauber} a further approximation  is suggested  arguing that
if the form factor of the deuteron decreases much more rapidly than the
scattering amplitude as function of $q'$, then the integral
in Eq.~(\ref{eq3.34}) may
be approximated by
\begin{equation}
\delta \sigma_{tot}  =\frac{2}{k^2}  m^2 ( 2 \pi)^5 \;
\Re e \left[t_s^2( p,x=1)\right] \int_0^{q_{max}} d q' \; q' S(q') ,
\label{eq3.35}
\end{equation}
where the scattering amplitude in forward direction ($x=1 $ or
$\theta=0$) is taken out of the integral.

\section{Results and Discussion}

Our explicit calculations are based on an interaction chosen as
superposition of two Yukawa interactions of the Malfliet-Tjon~\cite{malfliet}
 type,
\begin{equation}
V({\mathbf{p}}', {\mathbf{p}})=\frac{1}{2\pi^{2}}
\left(
\frac{V_{R}}{({\mathbf{p}}'-{\mathbf{p}})^{2}+\mu^{2}_{R}} -
\frac{V_{A}}{({\mathbf{p}}'-{\mathbf{p}})^{2}+\mu^{2}_{A}}
\right),
\label{eq4.1}
\end{equation}
in which the parameters, given in Table~\ref{table1},
 are fitted such that a bound state, the `deuteron'
is supported with a binding energy $E_d = -2.23$~MeV.
This interaction enters the nonrelativistic Faddeev equation for identical
bosons via the symmetrized two-body $t$-matrix.
The Faddeev equation is exactly solved without a partial wave
decomposition, using momentum vectors and angles between
them. This allows us to calculate three-body observables in the GeV region.
The details of the calculations are given in Ref.~\cite{Liu:2004tv}.
The deuteron wave function entering the form factor of
Eq.~(\ref{deutformf}) is obtained as solution of a bound state equation
with the potential of Eq.~(\ref{eq4.1}) as input.
As a note,
we are of course aware that this is a  model study, in which the
two-body $t$-operators are generated through a non-relativistic
Lippmann Schwinger equation from a simple Hermitian two-body force.

\subsection{Total Cross Sections}

\begin{figure}
\begin{center}
 \includegraphics[width=13cm]{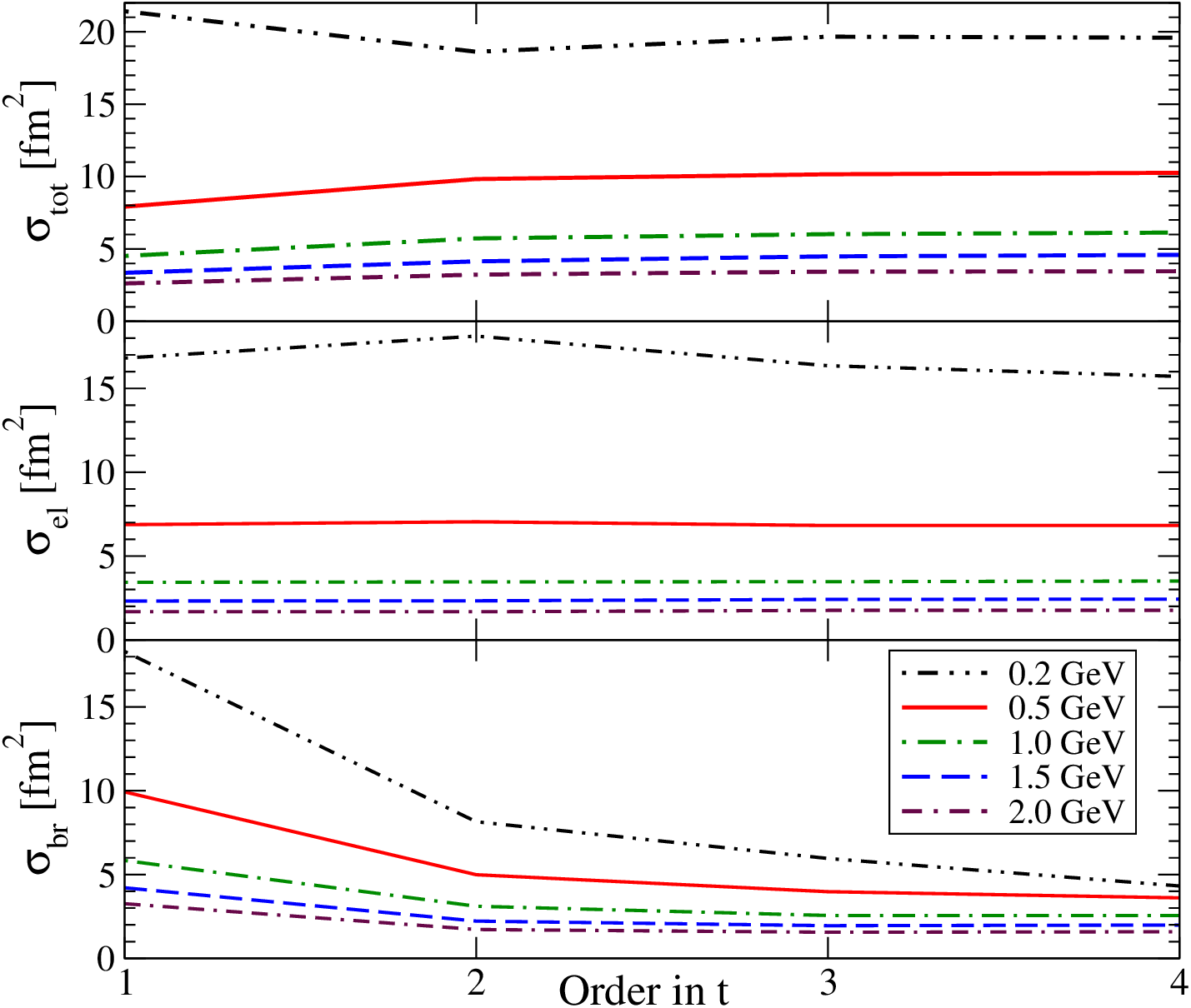}
\end{center}
\caption{(Color online) The total cross section (top panel), the
elastic total (middle panel) and the breakup total cross section
(bottom panel) as function of the order $t$ in the multiple scattering
series for selected laboratory projectile energies indicated in the
figure. Starting from the first order in the Faddeev calculation, the
next three higher orders (rescattering terms) are successively added.
\label{fig1}}
\end{figure}

First we want to explore the convergence of the Faddeev multiple scattering
series when considering total cross sections.
The top panel of
Fig.~\ref{fig1} shows the total cross section for three-body scattering
calculated using Eq.~(\ref{eq:2.7}) as function of the order in the two-body
$t$-matrix when adding up successively the terms of the multiple scattering
series created by Eq.~(\ref{eq:2.1}). The convergence is shown for selected
projectile laboratory energies ranging from 200~MeV to 2~GeV. For energies
larger than 1~GeV the curves are essentially flat, meaning that the first
order term given by $T=tP$ is already sufficient to capture the result of a
full Faddeev calculation. At 500~MeV one needs at least one
iteration (or one rescattering term) to reach the exact Faddeev result,
whereas at 200~MeV one needs more terms, which however give small
contributions.
The middle part of the panel shows the total cross section for elastic
scattering and the bottom one the total cross section for breakup
reactions. While the total cross section for elastic scattering converges
very fast for projectile energies above 500~MeV, the one for breakup
reactions requires at least one rescattering term even in the GeV range.

\begin{figure}
\begin{center}
 \includegraphics[width=12cm]{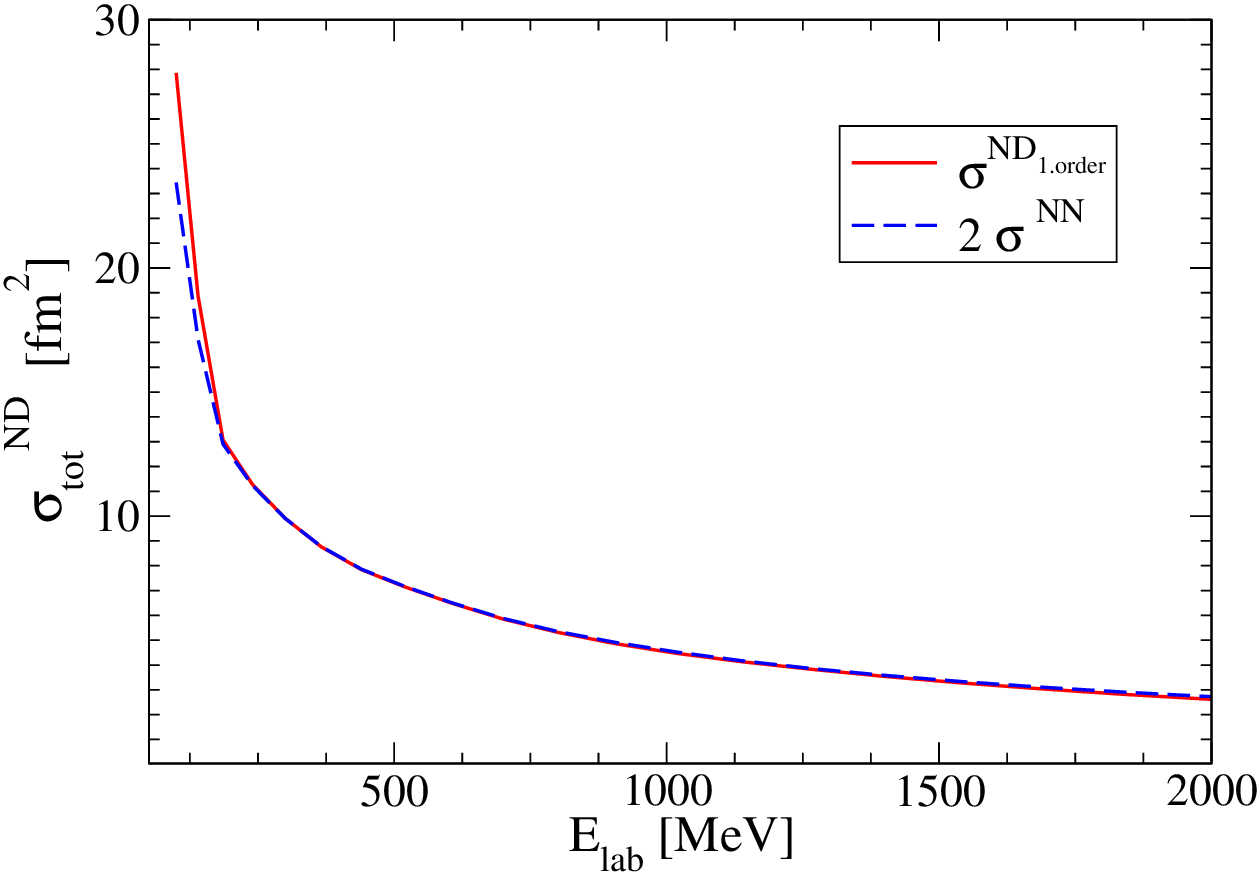}
\end{center}
\caption{(Color online) The total cross section for three-body
scattering as function of the laboratory projectile energy. Shown
are the first order (in $t$) Faddeev calculation and twice the
two-body total cross section $\sigma^{\rm NN}$ corresponding to the
first order term in a Glauber calculation as well as in the expansion of
Ref.~\cite{Elster:1998ry}.
\label{fig2}}
\end{figure}

Our Glauber calculation contains only a first order term and a rescattering
correction. First, we concentrate on the total cross section and
compare each order separately with the corresponding
order of the Faddeev calculation. The first order terms are compared in
Fig.~\ref{fig2} as function of the projectile laboratory energy. According to
Eq.~(\ref{eq3.10}) the first order contribution to the total $n+d$ cross
section is simply given by twice the two-body total cross section. From
Fig.~\ref{fig2} we see that for energies greater than 200~MeV both first
order terms agree. This is not surprising, since the first order Faddeev term,
$T=tP$, contains twice the two-body t-matrix,
the permutation operator guarantees
scattering contributions from both constituents of the target.  Our Glauber
calculations start from the same two-body $t$-matrix as the Faddeev
calculation. From this point of view it is not surprising that both first order
calculations agree so well. The same conclusion was already reached in
Ref.~\cite{Elster:1998ry}.

\begin{figure}
\begin{center}
 \includegraphics[width=12cm]{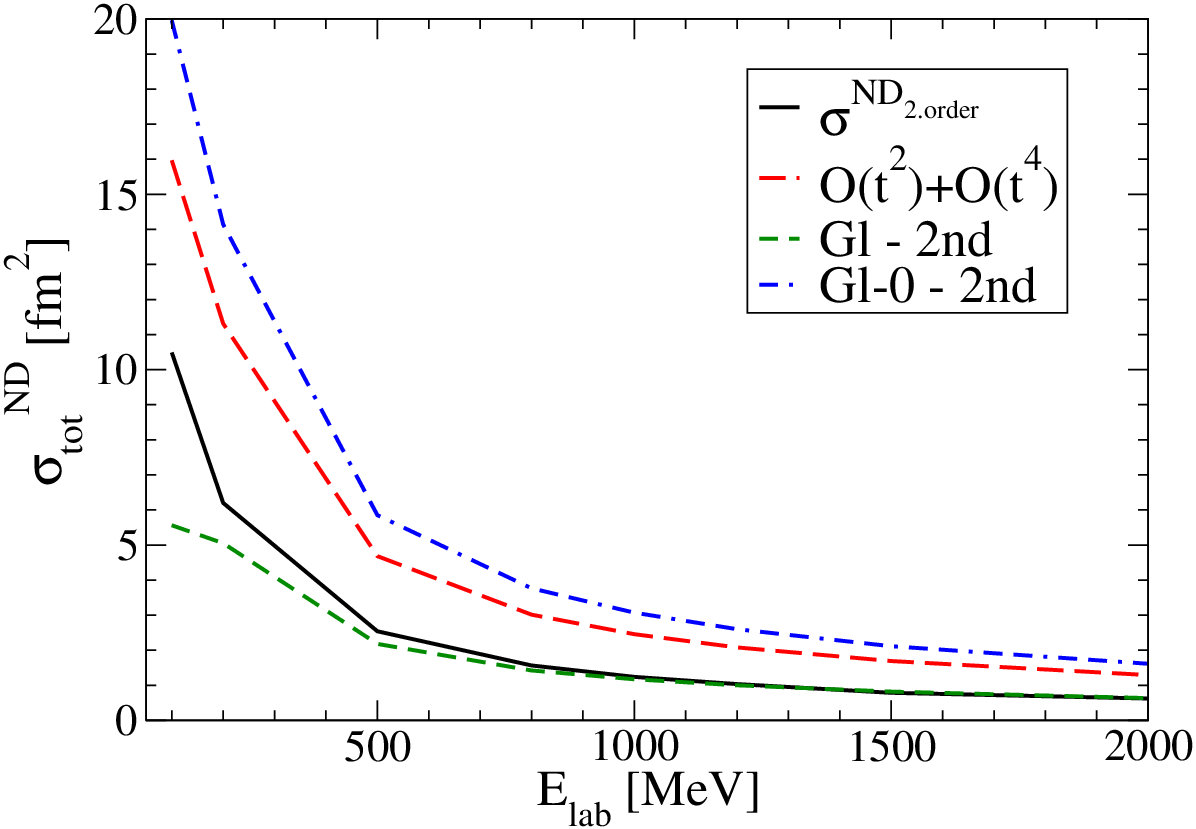}
\end{center}
\caption{(Color online) The second order correction to the total cross section for
three-body scattering as function of the laboratory projectile energy. The solid line
represents the second order term, $tPG_0tP$, of the Faddeev multiple scattering
series and the long dashed line the correction obtained in Ref.~\cite{Elster:1998ry}
from the high energy limit of the 2$^{nd}$ order Faddeev term. The short dashed line
corresponds to the 2$^{nd}$ order Glauber correction $\delta \sigma_{tot}$ from
Eq.~(\ref{eq3.34}), while the dash-dotted line stands for the approximation of
Eq.~(\ref{eq3.35}).
\label{fig3}}
\end{figure}

A more crucial test is a comparison of the
second order correction term.  In Fig.~\ref{fig3} we show the first rescattering
correction to the $n+d$ total cross section, which corresponds to second order
contribution in $t$ of the Faddeev multiple scattering series.  The
Faddeev result is shown as solid line. A first observation is that
this rescattering term contributes significantly at lower energies, and is
still present at 2~GeV. Thus, the very naive expectation that the
$n+d$ total cross section comprises only the scattering from the target constituents
is not fulfilled in the energy regime up to 2~GeV which is considered here.
There is always a contribution due to rescattering.
The second order Glauber correction $\delta \sigma$
 from Eqs.~(\ref{eq3.13}) and (\ref{eq3.32}) is given by the short
dashed line. Here we see that in contrast to the first order contribution the
two lines only start to agree around 1~GeV. This may be understood when
having in mind that the second order Faddeev term is explicitly given as
$tP G_0 tP$, where the free propagator $G_0$ describes the arbitrary free
motion of the three particles in the intermediate state. The Glauber second order
term assumes that the target constituents are frozen. At sufficiently high
energy this assumptions seems to be sufficient to describe the rescattering
within the target, which sometimes is referred to as `shadowing' correction.

Franco and Glauber \cite{FrancoGlauber} suggest a further approximation to
$\delta \sigma$. They argue that if the deuteron form factor decreases much
more rapidly than the scattering amplitude as function of the integration
variable $q'$, then the integral of Eq.~(\ref{eq3.34}) may be approximated by
the expression of Eq.~(\ref{eq3.35}), where the scattering amplitude
is approximated by its value in forward direction and taken out of the integral.
The double-dash-dotted
curve of Fig.~\ref{fig3} represents this approximation.
It is quite obvious that this approximation is too simplistic and gives a
second
order contribution almost double the size of the original correction at all
energies considered. The long-dashed curve represents the analytic evaluation
of the second order Faddeev term in the optical theorem of Eq.~(\ref{eq:2.9}).
The contribution calculated in there overestimates
the second order contribution in a similar fashion. In
Ref.~\cite{Elster:1998ry}, integrals over
the two-body $t$-matrix and the deuteron wave function were evaluated using the
method of steepest descent. This is equivalent to extracting the dominant part
of an integral as a constant from the integral. The failure of
those two approximations should lead to the conclusion, that although it is
true that for higher energies the scattering amplitude is peaked in
forward direction~\cite{3dt}, the integration over the form factor is
important and should not be approximated further.

\begin{figure}
\begin{center}
 \includegraphics[width=12cm]{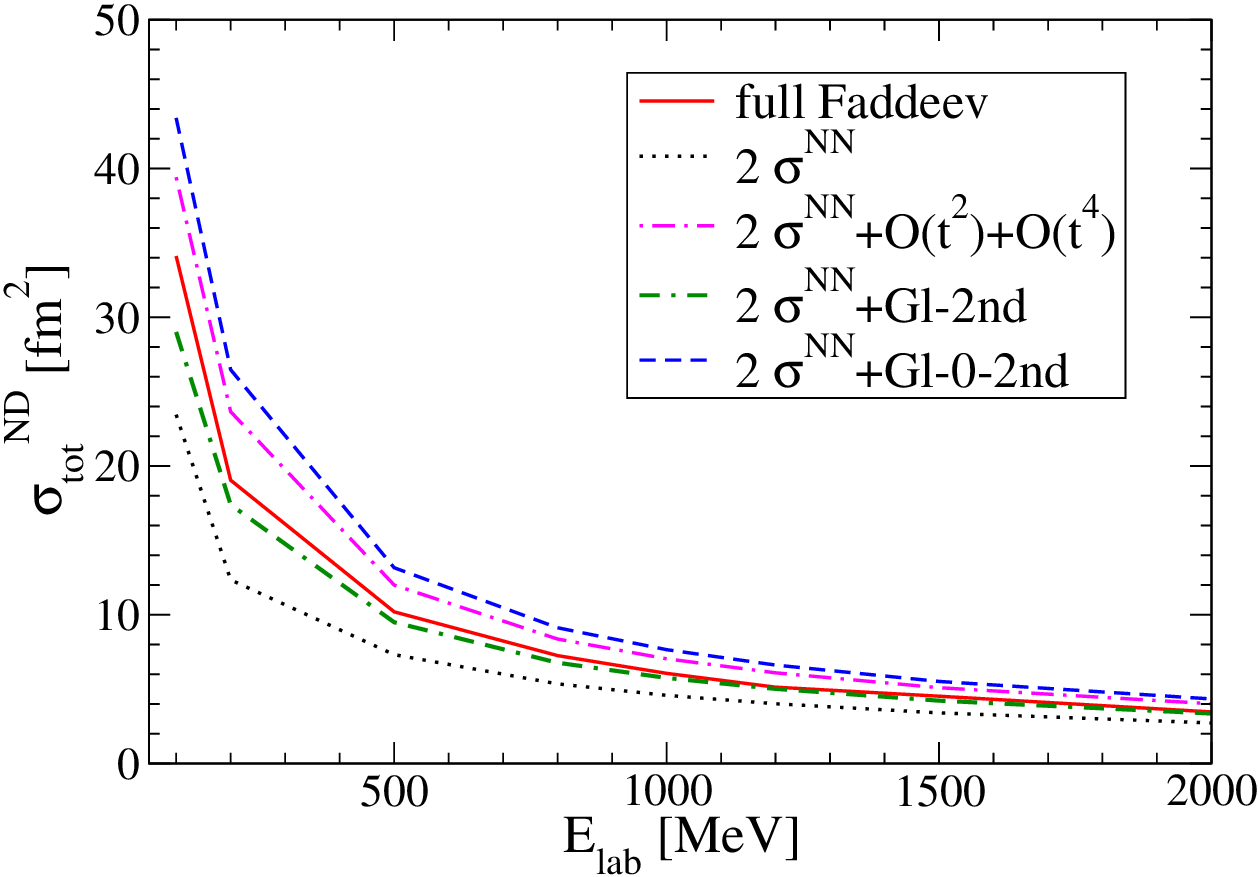}
\end{center}
\caption{(Color online) The total cross section for three-body scattering as function
of the laboratory projectile energy. The solid line represents fully converged
Faddeev calculations, the dotted line twice the two-body total cross section.
To this are added the  correction as extracted from the high energy
limit of the 2 $^{nd}$ order Faddeev term from Ref.~\cite{Elster:1998ry}
(dash-dotted
line), the second order Glauber correction
$\delta \sigma_{tot}$ (dash-dash-dotted line) and its approximation (dashed line).
\label{fig4}}
\end{figure}

After comparing the terms of the multiple scattering series separately, we show
in Fig.~\ref{fig4} the sum of the contributions as function of the projectile
laboratory energy. The solid line represents the exact Faddeev calculation, in
which the multiple scattering series is summed to all orders. As reference, the
dotted line represents twice the two-body total cross section $2 \sigma^{\rm NN}$,
to which the different second order contributions shown in Fig.~\ref{fig3} are
added. The Glauber calculations up to second order agrees with the full Faddeev
result starting from about 1~GeV. Both,  the simplification of the Glauber
second order term,  Eq.~(\ref{eq3.35}), and the high energy limit of
Eq.~(\ref{eq:2.9}) considerably overestimate the exact Faddeev result
at all energies considered.

\subsection{Differential Cross Sections}

Similar to the investigations of the total cross section, we want to start
comparing the first order calculations of the differential cross section at
various energies. The differential cross section for elastic $n+d$ scattering is
calculated from the operator $U$  for elastic scattering,
Eq.~(\ref{eq-U}), and given as function of the laboratory solid angle as
\begin{equation}
\frac{d\sigma^{\rm ND}}{d\Omega_l}= (2\pi)^4 \frac{2}{9}m^2
\frac{(x_l+\sqrt{x_l^2+3})^2}{\sqrt{x_l^2+3}} \left| \langle \phi_d {\hat q}
q_0\vert U \vert \phi_d {\bf q_0}\rangle \right|^2.
\label{eq4.2}
\end{equation}
Here $x_l=\cos \theta_l \equiv {\hat {\bf k'}}\cdot{\hat {\bf k}}$ represent
the laboratory scattering angle for the $n+d$ system,
and ${\bf q_0}$ its c.m. momentum.
For calculating the Glauber differential cross section from Eq.~(\ref{eq3.11}),
we have to consider the projection of the momentum transfer ${\bf q}$ on to a
plane either perpendicular to the laboratory momentum ${\bf k}$ or to the
sum of incoming and outgoing momenta ${\bf k}+{\bf k'}$, as indicated in
Eqs.~(\ref{eq3.24}) and (\ref{eq3.25}). Since both choices are employed in the
literature, we will consider both in this study.

\begin{figure}
\begin{center}
 \includegraphics[width=12cm]{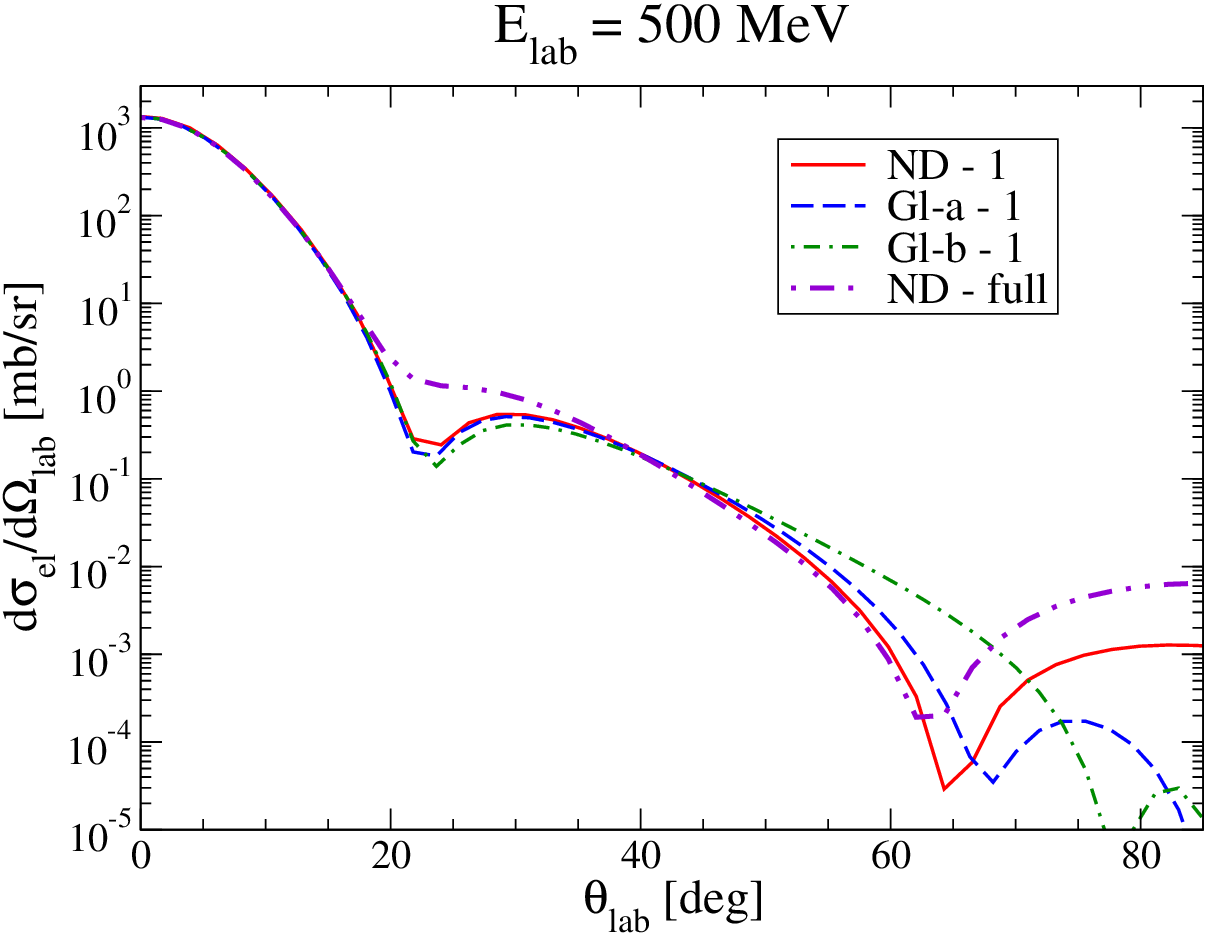}
\end{center}
\caption{(Color online) The differential cross section for elastic three-body
scattering at 500~MeV laboratory projectile energy
 as function of the laboratory scattering angle. The solid line represent the
first order Faddeev calculation. The dashed and dash-dotted line stand for the
first order Glauber calculations for the two choices (a) and (b) of the
direction of the unit vector $\hat n$ (see text). For comparison, the
dash-double-dotted line represents the full Faddeev calculation.
\label{fig5}}
\end{figure}

Taking the considerations from the previous subsection as guidance, we
start our comparison at 500~MeV projectile energy and show in Fig.~\ref{fig5}
the first order calculations for the differential cross section
as function of the laboratory scattering angle. The solid line
represents the first order Faddeev calculation, and the dashed and dash-dotted
lines the first order Glauber calculations with the choices (a),
Eq.~(\ref{eq3.24}),  and (b), Eq.~(\ref{eq3.25}), for the
projection of the momentum transfer. Both choices lead to identical results
for small scattering angles. The first small deviation can be seen
around the first diffraction minimum. Though the Glauber approximation should
only be valid in forward direction, we plot the entire allowed angle region in
order to obtain quantitative insights in the region of angular validity.
Both Glauber calculations start to deviate from the Faddeev first order
calculation  as well as
from each other at $\theta_l \approx 50^o$, which corresponds to a momentum
transfer of about 740~MeV. For comparison, we also add the fully converged
Faddeev result as dash-double-dotted line to the figure.
An obvious difference between the first order and the fully converged
Faddeev calculations is the first diffraction minimum, which is filled in
by rescattering corrections. Rescattering contributions are also important
for the large angle (high momentum transfer) behavior of the differential cross
section~\cite{Liu:2004tv,Lin:2008sy}.

\begin{figure}
\begin{center}
 \includegraphics[width=12cm]{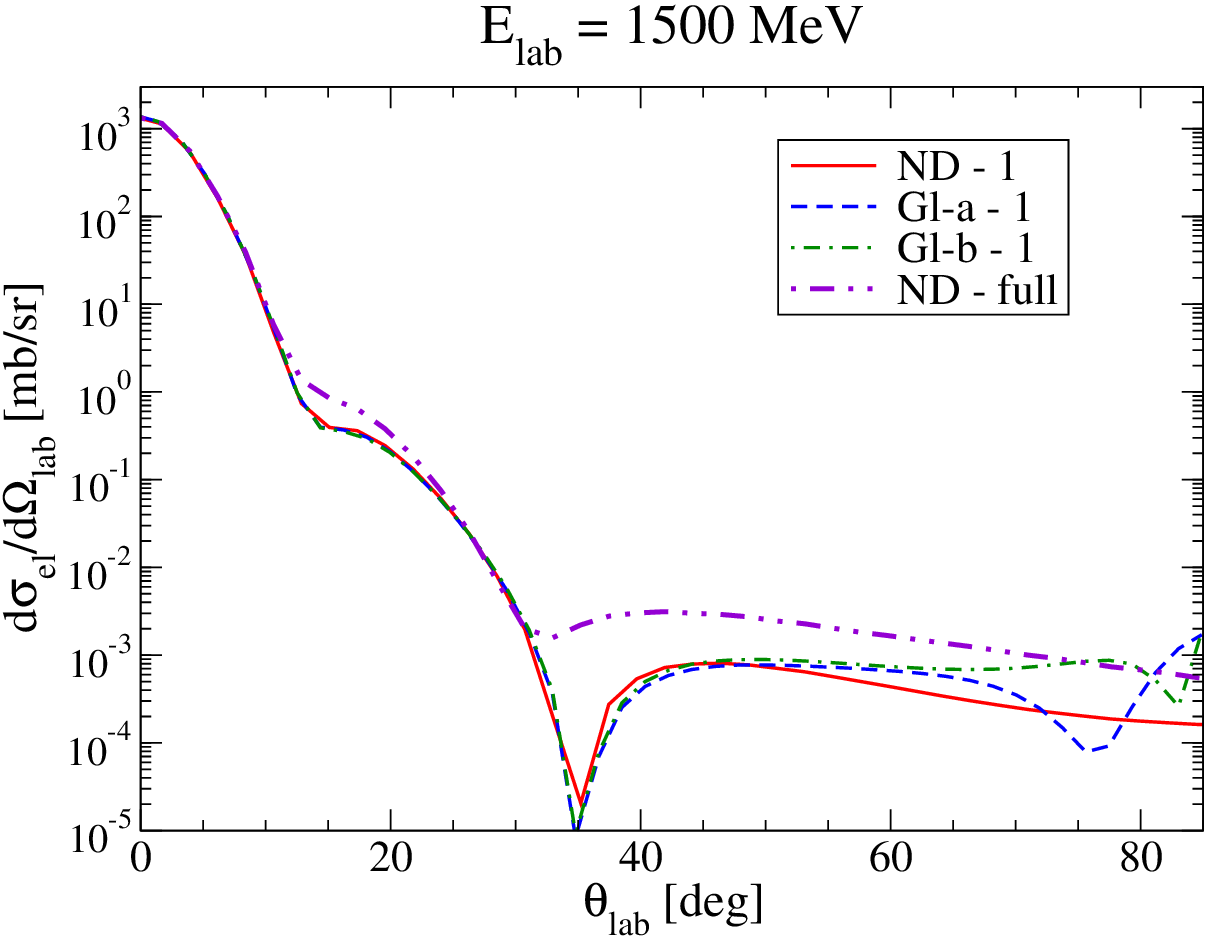}
\end{center}
\caption{(Color online) The differential cross section for elastic three-body
scattering at 1500~MeV laboratory projectile energy
 as function of the laboratory scattering angle. The meaning of the curves is
the same as in Fig.~\protect\ref{fig5}.
\label{fig6}}
\end{figure}

\begin{figure}
\begin{center}
 \includegraphics[width=12cm]{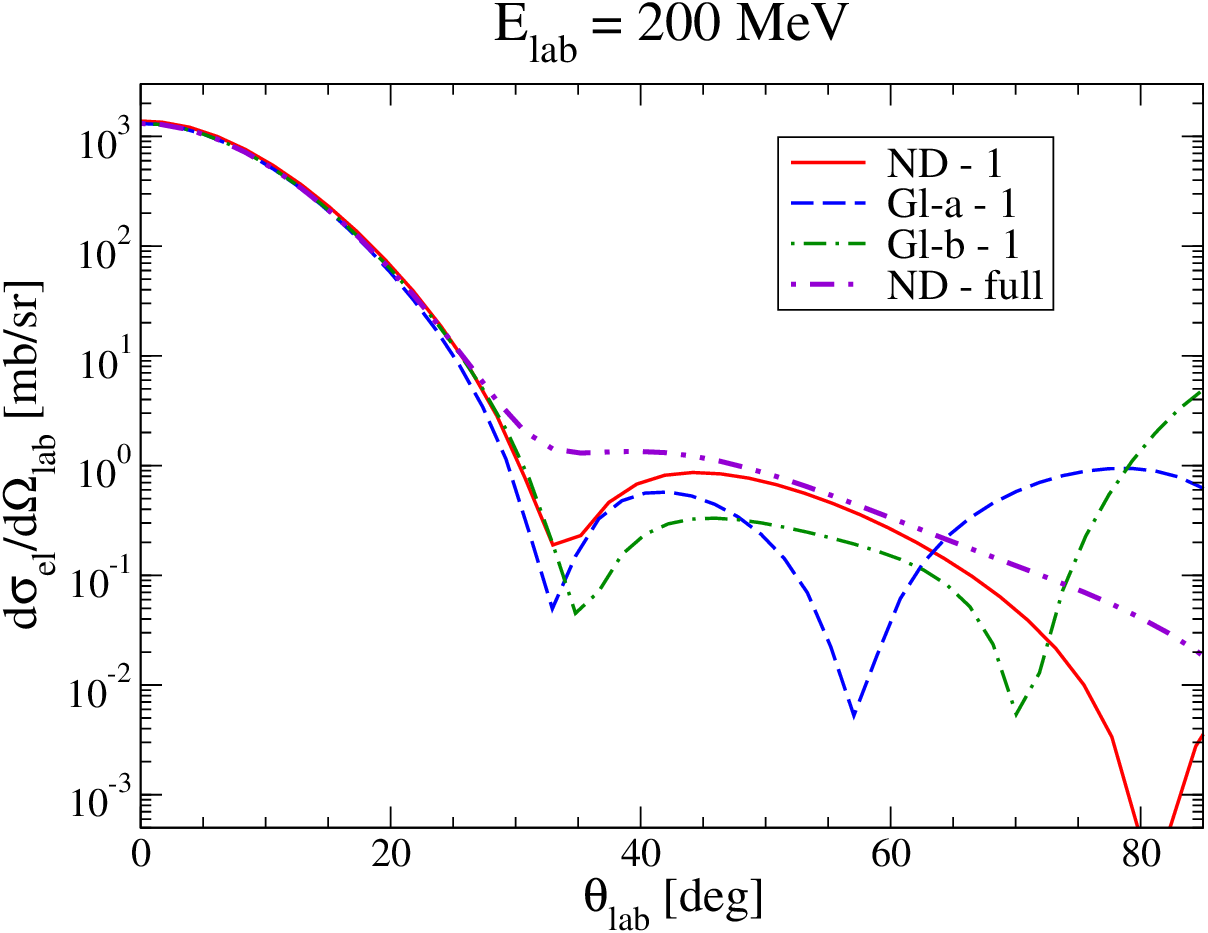}
\end{center}
\caption{(Color online) The differential cross section for elastic three-body
scattering at 200~MeV laboratory projectile energy
 as function of the laboratory scattering angle. The meaning of the curves is
the same as in Fig.~\protect\ref{fig5}.
\label{fig7}}
\end{figure}

\begin{figure}
\begin{center}
 \includegraphics[width=12cm]{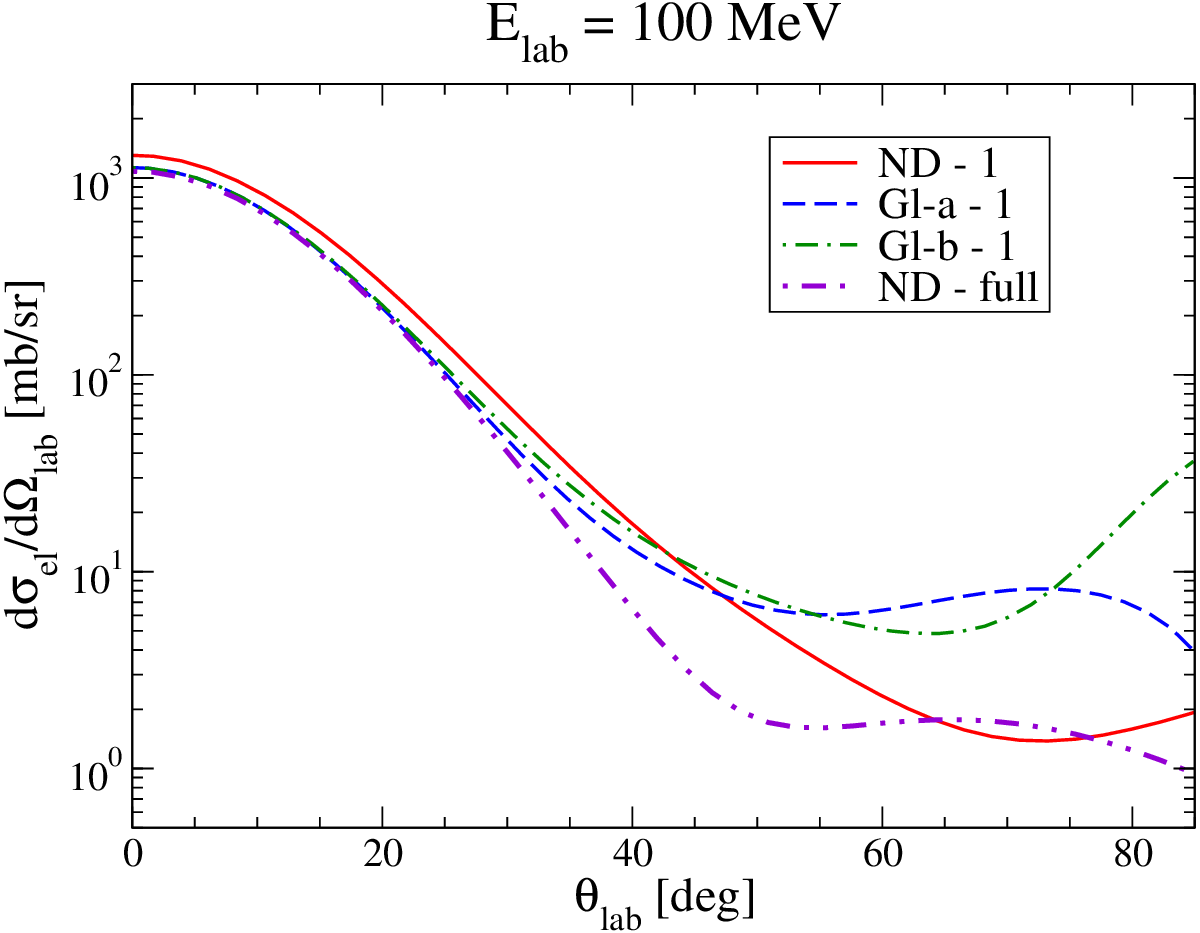}
\end{center}
\caption{(Color online) The differential cross section for elastic three-body
scattering at 100~MeV laboratory projectile energy
 as function of the laboratory scattering angle. The meaning of the curves is
the same as in Fig.~\protect\ref{fig5}.
\label{fig8}}
\end{figure}

Fig.~\ref{fig6} shows the differential cross section for similar
calculations at
the considerably higher energy of 1.5~GeV. Again, the first order Faddeev
calculation and the first order Glauber calculations with the two different
choices of ${\bf q}_\perp$ are close to each other up to $\theta_l \approx
50^o$, which now corresponds to a momentum transfer of about 1250~MeV.
However, the first order calculations differ from the fully converged
Faddeev calculation already at much smaller angles, around $\theta_l \approx
12^o$, which corresponds to roughly 350~MeV in the momentum transfer.
While the Glauber formulation is intended as high energy approximation,
we are interested in
exploring how the differential cross sections
of the first order Glauber and Faddeev calculations compare at lower energies.
In Figs.~\ref{fig7} and \ref{fig8} the differential cross sections for
laboratory projectile energies of 200 and 100~MeV are displayed.
For 200~MeV the two different choices of ${\bf q}_\perp$ in the Glauber first
order calculations already differ at $\theta_l \approx 25^o$, corresponding to
a momentum transfer of about 270~MeV. At roughly the same angle the
first order Faddeev calculations also deviates from the converged solution.
It is well known that the Faddeev multiple scattering series converges very
slowly at such low energies~\cite{wgphysrep}, so this finding is
not surprising. However, at very small angles,
all calculations shown roughly agree. This becomes different, if one looks at
the even lower energy of 100~MeV laboratory projectile energy.
Here the first order Glauber calculations do not agree with the first
order Faddeev calculation even at small angles. It is also very obvious that at
this low energy the first order Faddeev calculations is quite different from
the fully converged one. It is however noteworthy, that both first order
Glauber calculations are closer to the converged Faddeev result in forward
direction than is the calculation based on the first order Faddeev term.
The two different choices
for ${\bf q}_\perp$ start to differ around $\theta_l \approx 25^o$, which
corresponds  now only to a momentum transfer of roughly 180~MeV.
The overall observation
is that at such a low energy neither a first order Faddeev calculation nor a
Glauber calculation are a good representation of the fully converged Faddeev result.

\begin{figure}
\begin{center}
 \includegraphics[width=13cm]{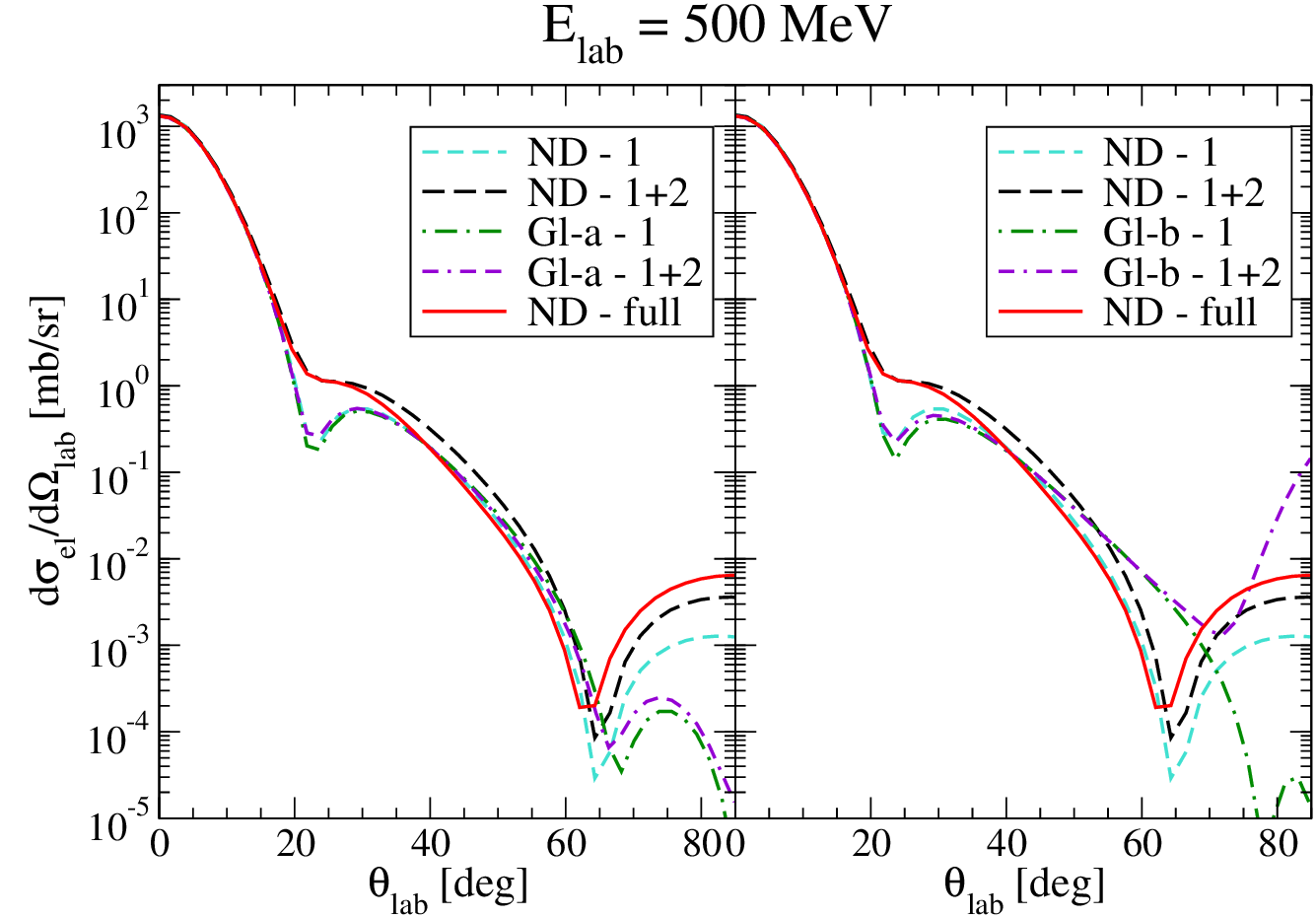}
\end{center}
\caption{(Color online) The differential cross section for elastic
three-body scattering at 500~MeV laboratory projectile energy
as function of the laboratory scattering angle.
In both panels short dashed line represents the first order Faddeev
calculation, the long dashed line the first and second order are considered,
and the solid line stands for the full Faddeev calculation.
The left panel shows the first order (dash-dotted line) and first plus second
order (double-dash-dotted line) Glauber calculation for the choice (a)
of the direction of the unit vector $\hat n$ (see text).
The right panel contains the same for the choice (b).
\label{fig9}}
\end{figure}

Next we add the Glauber rescattering correction of Eq.~(\ref{eq3.14}) to the
differential cross section. In Fig~\ref{fig9} we show the effect of this
term on the differential cross section at 500~MeV laboratory projectile
energy together with the contribution of the second order Faddeev rescattering
term. While the Faddeev first order rescattering term gives a large
contribution to the first minimum, the Glauber rescattering term fails to do
this, independent of the choice of the unit vector ${\hat n}$. Obviously,
the minima are more sensitive to the interference of the two terms of the
scattering amplitude. Thus here the very different structure of the
Faddeev and Glauber rescattering terms are clearly visible.
It should also be noted that in the first minimum one Faddeev rescattering
term is already sufficient to coincide with the fully converged Faddeev
calculation. At slightly larger angles more rescattering terms are required
in the Faddeev calculation to reach convergence.

\begin{figure}
\begin{center}
 \includegraphics[width=13cm]{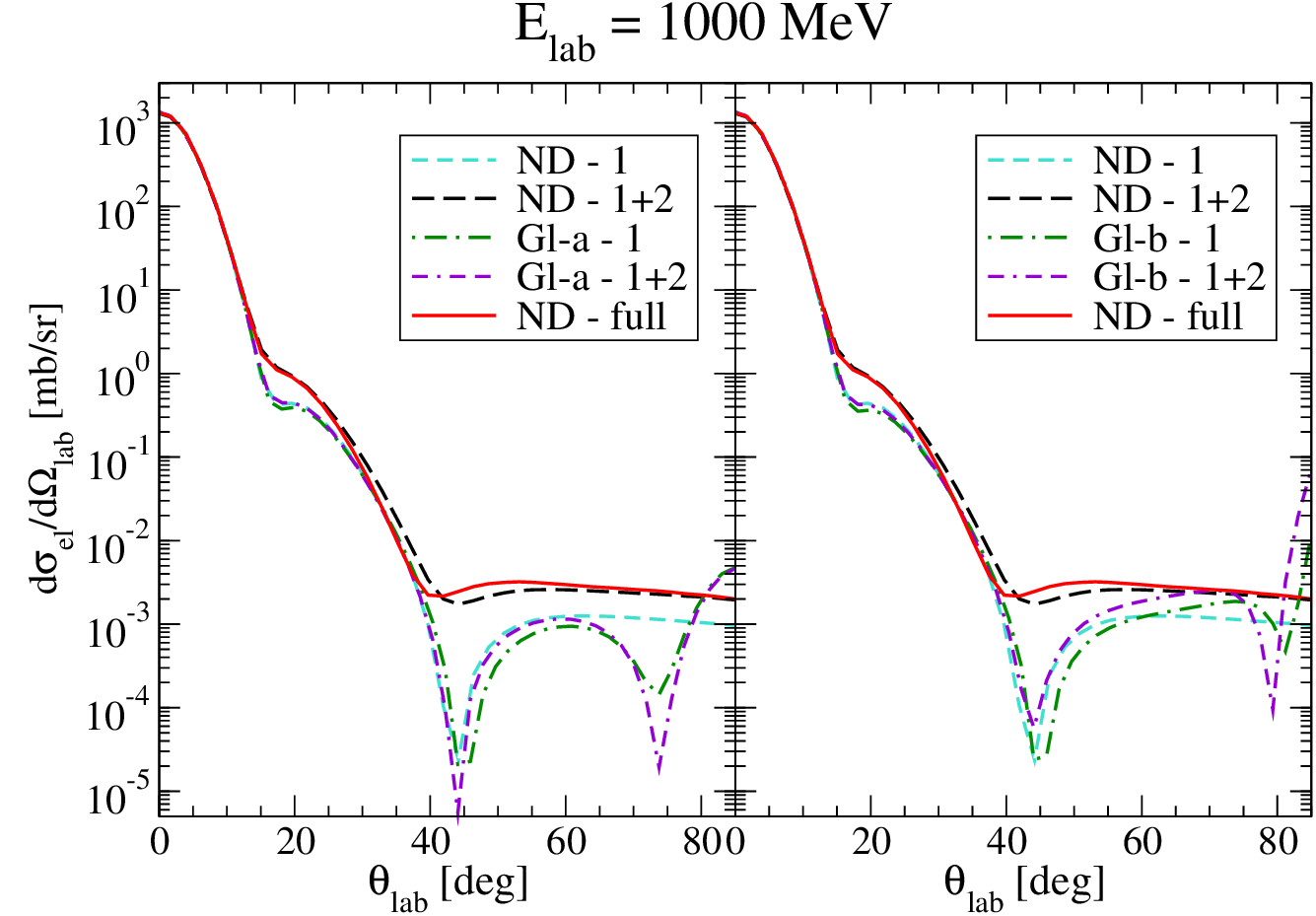}
\end{center}
\caption{(Color online) The differential cross section for elastic
three-body scattering at 1000~MeV laboratory projectile energy
as function of the laboratory scattering angle.
The curves have the same meaning as in Fig.~\protect\ref{fig9}.
\label{fig10}}
\end{figure}

\begin{figure}
\begin{center}
 \includegraphics[width=13cm]{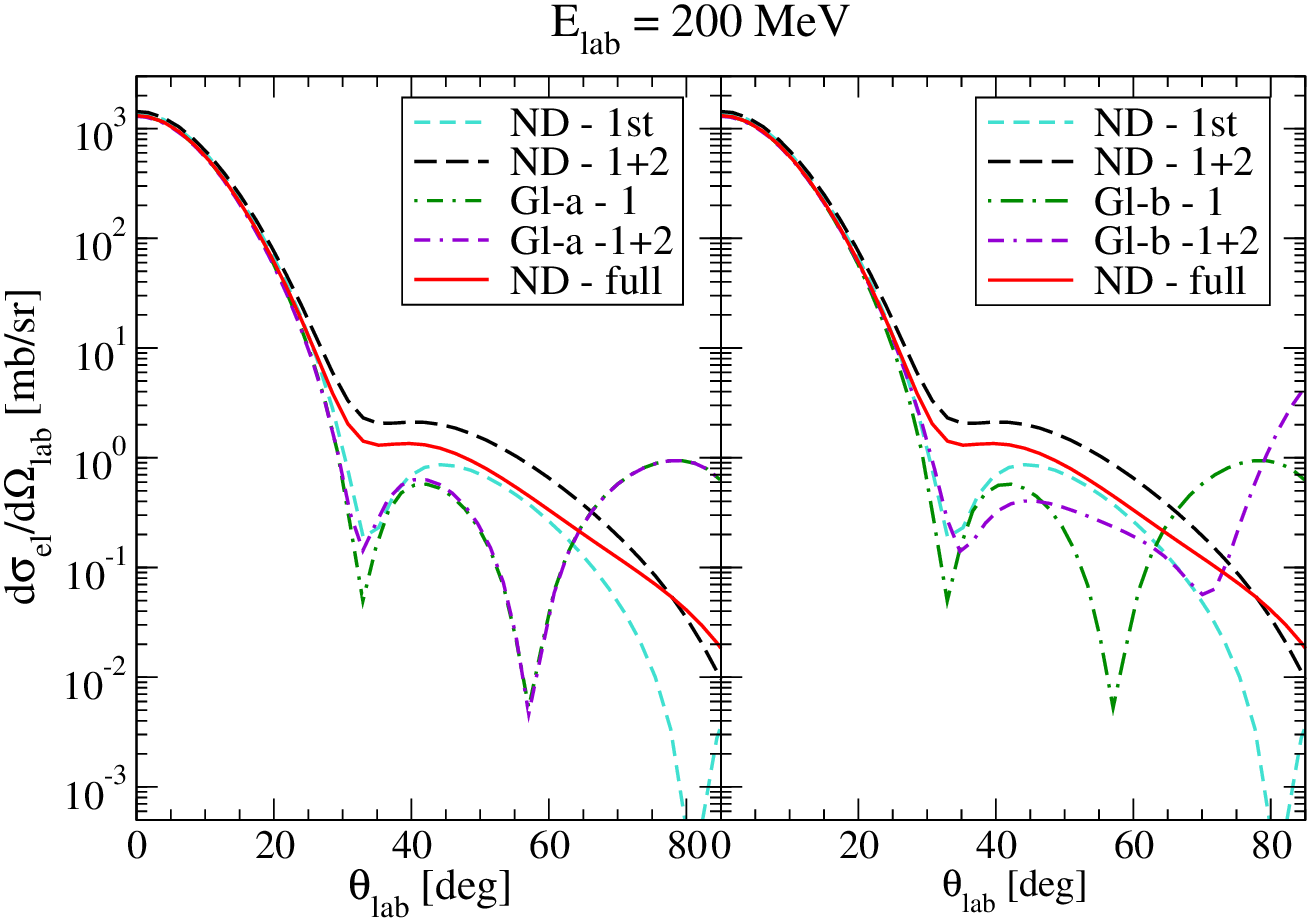}
\end{center}
\caption{(Color online) The differential cross section for elastic
three-body scattering at 200~MeV laboratory projectile energy
as function of the laboratory scattering angle.
The curves have the same meaning as in Fig.~\protect\ref{fig9}.
\label{fig11}}
\end{figure}

This general feature of the difference between the two approaches does not
change with increasing energy. In Fig~\ref{fig10} the differential cross
section is shown for a laboratory projectile energy of 1~GeV. While the total
cross section for both, Faddeev and Glauber calculations started to agree
at energies of 1~GeV and higher when adding the second order correction
term in both schemes, this is not the case for the differential
cross section. The second order Faddeev rescattering term again  gives a
much larger contribution in the first minimum than the Glauber rescattering
correction. This trend continues up to 2~GeV, the highest energy we consider
here. We also can observe that with increasing energy the difference between
the two choices of ${\hat n}$ decreases. To investigate at which energies
there are differences, we need to consider lower projectile energies.
In Fig.~\ref{fig11} the differential cross section for 200~MeV projectile
laboratory energy is shown. Here we can first observe, that the second order
Glauber correction also increases the cross section in the first minimum,
but again by far less than the Faddeev rescattering term does. We also
observe a distinct difference between the two choices of ${\hat n}$.
At 200~MeV the choice (a) almost has no effect at larger angles, whereas
choice (b) shows a big contribution. However, since the Glauber expression
should be valid only in forward direction, one should not put any
physical relevance to this difference.

\begin{figure}
\begin{center}
 \includegraphics[width=14cm]{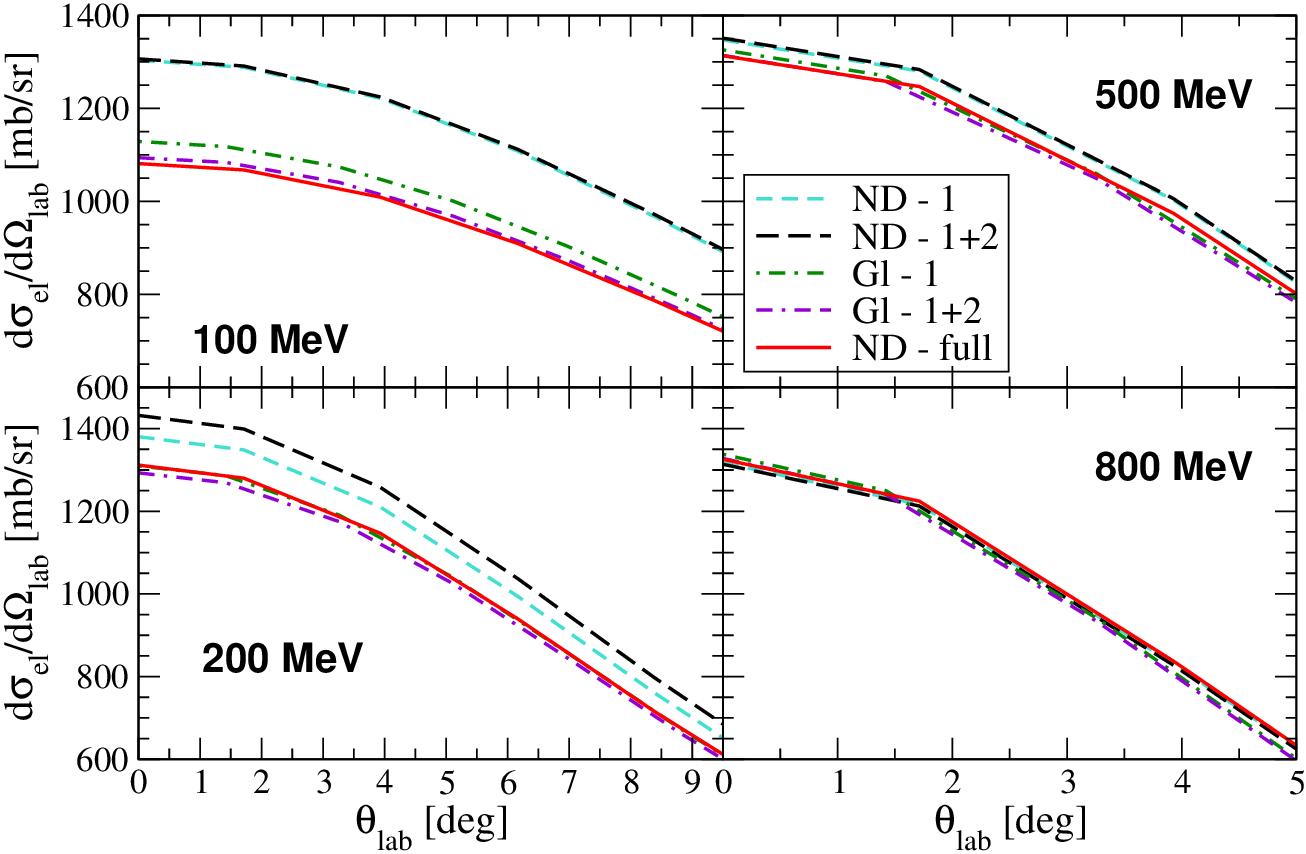}
\end{center}
\caption{(Color online) The differential cross section for elastic three-body
scattering in forward direction for 100~MeV (left upper panel), 200~MeV (left
lower panel), 500~MeV (right upper panel), and 800~MeV (right lower panel)
laboratory projectile energy.
The short dashed line represents the first order Faddeev
calculation, the long dashed line the first and second order are considered,
and the solid line stands for the full Faddeev calculation.
The first order and first plus second
order Glauber calculation for the choice (b)
of the direction of the unit vector $\hat n$ are given by the dash-dotted and
double-dash-dotted lines.
\label{fig12}}
\end{figure}

After having seen that the Glauber rescattering term does not compare well
with the corresponding Faddeev term in diffraction minima, we want to
consider the differential cross section in the very forward direction. This
is the physical regime for which the Glauber ansatz was developed and where
it should perform best.
In Fig.~\ref{fig12} we show the three-body differential cross section for
very forward angles for laboratory projectile energies of 100~MeV, 200~MeV,
500~MeV, and 800~MeV. We again compare the results from the first and second
order terms in the Faddeev multiple scattering series to the fully converged
Faddeev calculation, and the first order Glauber calculation with its
rescattering correction to the Faddeev calculations. Both choices of
${\hat n}$ coincide for very small angles. Thus we arbitrarily choose
(b) for the Glauber calculations of Fig.~\ref{fig12}. Since we already
established from the comparison of the total cross sections, that the Glauber
calculations do very well for energies of 1~GeV and higher, we concentrate
here on low and intermediate range energies. At 800~MeV essentially all
calculations agree with each other in the very forward direction. At 500~MeV
the first order Faddeev term slightly over-predicts the fully converged
result. The first Faddeev rescattering term has almost no effect, and one
needs higher order rescattering contributions to achieve a converged result.
 The Glauber first order calculation differs from the
Faddeev first order and only slightly over-predicts the converged Faddeev
result. The contribution of the Glauber rescattering term is small, and
lowers the cross section for the very forward angles directly to the full
Faddeev result. The situation at 200~MeV is similar. Again, the
first order Faddeev and Glauber calculations differ, and the Glauber
calculation already coincides with the fully converged Faddeev result.
The Glauber rescattering contribution again slightly lowers the
cross section in forward direction. At this energy it is interesting to
observe that the first rescattering contribution of the Faddeev multiple
scattering series increases the cross section in forward direction.
To reach convergence eight terms of the multiple
scattering series must be summed up at this energy.
At 100~MeV the Glauber first order calculation is
already quite close to the fully converged Faddeev results, in contrast to
the first order Faddeev calculation. The Glauber rescattering correction then
lowers the cross section essentially to the exact Faddeev result.
Thus, for the differential cross section in very forward direction the
Glauber calculation corrected by the rescattering term captures the
exact Faddeev result already at 100~MeV. Fig.~\ref{fig8} shows that
this good agreement is valid to about $\theta_l \approx 20^o$.

The behavior of the differential cross section in the extreme forward
direction appears in a very similar fashion in  the total
cross section for elastic scattering, $\sigma_{el}$,
which is obtained by integrating over the
differential cross section. Table~\ref{table2} lists
the total cross sections for elastic scattering as function of the laboratory
projectile energy for the converged Faddeev calculations, Faddeev calculations
including only the first two rescattering terms as well as the Glauber
calculations. We see that already from 200~MeV on the first order Glauber
calculations match the converged Faddeev results reasonably well. The
Glauber rescattering correction is quite small, and always lowers the cross
section, whereas the  first Faddeev rescattering correction increases  the
cross section, and only the second rescattering correction (3rd order in the
multiple scattering series) lowers it. It is also obvious from the table that
at 100~MeV many more terms of the Faddeev multiple scattering series are
necessary to obtain a converged total cross section. The Glauber result with
rescattering correction is similar in size to the Faddeev result with two
rescattering corrections, though this is most likely accidental.
A Glauber calculation for the total cross section for elastic scattering at
100~MeV is not trustworthy anymore, since the differential cross sections
do not really agree.

The total cross section is the sum of the total cross sections for elastic
scattering and breakup reactions, $\sigma^{\rm ND}_{tot} = \sigma^{\rm
ND}_{el} + \sigma^{\rm ND}_{br}$. In Refs.~\cite{Liu:2004tv,Lin:2008sy}
this relation was used to estimate the quality of the numerical solutions
of the Faddeev equation. In a Faddeev calculation the total cross section for
breakup reactions is obtained by integration over the five-fold differential
cross section for breakup, i.e. by summing over all possible breakup
configurations.
Here we can employ this relation to obtain a total cross section
for breakup reactions within the Glauber framework.  This is done in
Table~\ref{table3}. The left side lists the Faddeev total cross sections for
elastic scattering and breakup reactions with their sum, the total cross
section, as function of laboratory projectile energy. The right side lists
the Glauber total cross section obtained from Eq.~(\ref{eq3.12}). The Glauber
total cross
sections for breakup reactions are obtained by subtracting the Glauber total cross
sections for elastic scattering from Table~\ref{table2} from these numbers.
For energies less than 500~MeV the so obtained breakup cross sections
are definitely not competitive with the Faddeev results.
However, at 1.5~GeV, the Glauber breakup cross section starts to match
the Faddeev breakup cross section within 5\%.
Though this finding may look surprising, it could indicate that at sufficiently
high energy the dominant
ejectiles of a breakup reaction exit into the forward cone.
That region of phase space however is the one, for which the Glauber ansatz
was designed. Thus it seems that for GeV projectile energies obtaining
a total cross section for breakup reactions from the Glauber total and
elastic cross section via the optical theorem captures the overall breakup
reaction. Of course the description of detailed
exclusive $n+d$ observables
will never be accessible in the Glauber formulation.

\section{Summary and Conclusions}

In this study we perform fully converged Faddeev calculations for three
identical bosons (our model for the $n+d$ system)
in the intermediate energy regime between 100~MeV and 2~GeV. We
calculate total cross sections as well as differential cross sections for
elastic scattering.  The key point of those calculations is the use
of vector momenta in the formulation, so that all partial waves are
automatically included.
We then calculated the same cross sections using Glauber
formulation~\cite{Harr64,FrancoGlauber,BJM}. However a key difference to this
early work is that the two-body scattering amplitude
entering our Glauber expressions  is the solution of a two-body
Lippmann-Schwinger equation, i.e. the same input as is used in the Faddeev
calculations. As two-body interaction we employ a superposition of
two Yukawa  terms, one attractive, the other repulsive, for which the
parameters are chosen such that a bound state at the empirical deuteron binding
energy is supported. The deuteron wave function and form factor are also
calculated from this interaction.

By using the identical two-body input for the Faddeev and Glauber calculations we
make sure that both calculations are based on the same ingredients.
This way we can clearly attribute differences in the observables to the
different treatment of multiple scattering in the two formulations.
The first order term in the Glauber formulation is given by twice the
two-body scattering amplitude,
which is folded with the deuteron form factor to give
the differential cross section and leads to 2$\sigma^{\rm NN}$ for the total cross
section. The first order Faddeev term is given by $T=tP$, thus having a
similar structure. Comparing the total cross sections obtained from the first
order term leads to the conclusion that from projectile energies of about
200~MeV onward both calculations agree. For the differential cross section
agreement to about $\theta_l \approx 45^o$ is achieved for projectile energies of
500~MeV and higher. For 200~MeV the angular range of agreement is already less,
and at 100~MeV the first order calculations do not agree with each other.
In general we can conclude, that in first order the Glauber and Faddeev
calculations for observables considered, total and differential cross sections
for elastic scattering are remarkably close for energies higher than 200~MeV
projectile energy.

The consideration of the first rescattering correction in both formulation
shows that there is a distinct dependence on the observables considered.
As a reminder, the Glauber formulation leads only to one rescattering
correction, whereas the Faddeev formulation as integral equation has an in
principle infinite series of rescattering corrections. As far as the
correction to the total cross
section is concerned, the contribution from the
first Glauber and Faddeev rescattering term
starts to become close at 500 MeV and is identical in size at 1~GeV. We also
want to point out,
that this rescattering correction to the total cross section, the
shadowing (better named anti-shadowing here, since the contribution
increases the cross section), is still present at 2~GeV, the highest energy we
considered. Thus we conclude, that for a good description of the total cross
section in the intermediate energy regime the consideration of
the second order correction in the Glauber as well as in the Faddeev formulation
is essential.

For the differential cross section for elastic scattering the conclusions are
twofold.
First, when considering  the very forward
direction, we find that at projectile energies of 100 and 200~MeV
the second order correction
brings the Glauber calculation of the differential cross section in quite good
agreement with the result of the fully converged Faddeev calculation. In
contrast, the first Faddeev rescattering term even has the opposite effect and
moves the cross section
away from the converged calculation. At projectile energies of 500~MeV and
higher, the second order corrections in the very forward direction are extremely
small and the first order Faddeev and Glauber calculations already agree quite
well with each other and the converged Faddeev calculation.
A similar conclusion can be drawn when considering the total
cross sections for elastic scattering. Second, when
concentrating on the diffractive structure of the
differential cross section given in
our model, we see
distinct differences between the behavior of the Faddeev and Glauber second
order correction. Whereas the Faddeev correction fills in the minima, the
Glauber correction can not do that. A similar observation was made in
Ref.~\cite{Crespo:2007zza} in the context of proton elastic scattering from halo
nuclei considering energies from 100 to 200~MeV per nucleon. However we find
that even at much higher energies the Glauber second order correction does not come close
to the effect of the Faddeev correction. For energies higher than 500~MeV the
first Faddeev rescattering correction already captures the bulk of rescattering
corrections and coincides with the converged calculation in the first minimum.
Diffraction minima are well known to be sensitive to the dynamics of the system
or interference effects. The assumption of fixed target particles leading to the
specific form of the Glauber rescattering correction proves to be too simple
to capture the much more involved structure of the first Faddeev rescattering
term.
We also need to point out here, that
the diffractive structures shown in
the differential cross sections of this work are a result of our underlying
assumption of identical bosons. The true $n+d$ differential cross section has
only one minimum and one might speculate that a Glauber calculation at higher
energies including the rescattering term and being
based on nucleon-nucleon (NN) interactions describing the NN
observables has a chance to describe the $n+d$ total cross
and possibly the  differential cross section
certainly in forward direction quite well,
provided a relativistic formulation is going to be used.

\section*{Acknowledgments}
This work was performed in part under the
auspices of the U.~S.  Department of Energy, Office of
Nuclear Physics
under contract No. DE-FG02-93ER40756 with Ohio University, and
in part under NSF grant PHY-0653312  with the Ohio State University.

We thank the Ohio  Supercomputer Center (OSC) for the use of
their facilities under grant PHS206. In addition we
appreciate helpful comments from David R. Harrington.
Ch.E. thanks the nuclear
theory group of the Ohio State University for their warm
hospitality during the time this work was carried out.


\clearpage


\begin{table}
\begin{tabular}{|c|c|c|c|c|}
\hline
$V_{A}$ [MeV fm] & $\mu_{A}$ [${\mathrm{fm}}^{-1}$] & $V_{R}$[MeV fm]
  & $\mu_{R}$ [${\mathrm{fm}}^{-1}$] & $E_{d}$[MeV] \\ \hline
  626.8932 & 1.55 & 1438.7228 & 3.11 & -2.23  \\  \hline
\end{tabular}
\caption
{The parameters and deuteron binding energy for the Malfliet-Tjon
type
potential (MT3) of our calculation. As conversion factor
We use units such that $\hbar c$=197.3286~MeV fm~=~1.}
\label{table1}
\end{table}

\vspace{10mm}

\begin{table}
\begin{center}
\begin{tabular}{|c|ccc|c||cc|cc|} \hline
 & \multicolumn{4}{c|}{Faddeev} & \multicolumn{2}{c|}{Glauber (a)}
&\multicolumn{2}{c|}{Glauber (b)}\\
\hline
$E_{lab}$ [MeV] & $\sigma^1_{el}$ [fm$^2$] & $\sigma^{1+2}_{el}$ [fm$^2$] &
$\sigma^{1+2+3}_{el}$ [fm$^2$]& $\sigma_{el}^{\rm ND}$ [fm$^2$]&
 $\sigma^1_{el}$ [fm$^2$] & $\sigma^{1+2}_{el}$ [fm$^2$] & $\sigma^1_{el}$ [fm$^2$] &
$\sigma^{1+2}_{el}$ [fm$^2$]\\
\hline \hline
 100 &   40.66 & 44.47 &  30.27  & 26.53 &  29.58 &   28.24 & 32.60 & 30.83 \\
 200 &   16.81 & 18.12 &  15.84  & 15.31 &  15.32 &   15.06 & 15.74 & 15.46 \\
 500 &    6.87 &  7.04 &   6.82  &  6.74 &   6.59 &   6.52  &  6.63 &  6.56 \\
 800 &    4.29 &  4.33 &   4.34  &  4.32 &   4.23 &   4.19  &  4.25 &  4.21 \\
 1000&    3.43 &  3.46 &   3.47  &  3.46 &   3.44 &   3.41  &  3.45 &  3.43 \\
 1200&    2.91 &  2.92 &   2.95  &  2.96 &   2.90 &   2.88  &  2.91 &  2.89 \\
 1500&    2.32 &  2.33 &   2.41  &  2.41 &   2.35 &   2.33  &  2.35 &  2.34 \\
 2000&    1.68 &  1.68 &   1.76  &  1.76 &   1.78 &   1.77  &  1.78 &  1.77 \\
\hline \hline
\end{tabular}
\end{center}
\caption{The Total cross section for elastic scattering  for
the first 3 terms of the Faddeev multiple scattering series, the fully
converged Faddeev calculation, and the first two orders of Glauber
calculations based on choices (a) and (b) for the unit vector~${\hat n}$
The superscripts indicate the order in the two-body $t$ matrix.
} \label{table2}
\end{table}

\vspace{10mm}

\begin{table}
\begin{center}
\begin{tabular}{|c|ccc||ccc|} \hline
& \multicolumn{3}{c|}{Faddeev} & \multicolumn{3}{c|}{Glauber } \\
\hline
$E_{lab}$ [MeV] &$\sigma_{tot}^{\rm ND}$ [fm$^2$] & $\sigma_{el}^{\rm ND}$ [fm$^2$] &
$\sigma_{br}^{\rm ND}$ [fm$^2$]&
$\sigma_{tot}^{\rm Gl}$ [fm$^2$]& $\sigma_{br}^{\rm Gl}$(a)[fm$^2$] &
  $\sigma_{br}^{\rm Gl}$(b)[fm$^2$]\\
\hline \hline
 100 &   34.16 & 26.53 &  7.63  & 29.00 &  0.76 &  -   \\
 200 &   19.00 & 15.31 &  3.69  & 17.39 &  2.33 & 1.93 \\
 500 &   10.30 &  6.74 &  3.56  &  9.49 &  2.97 & 2.93 \\
 800 &    7.22 &  4.32 &  2.90  &  6.77 &  2.58 & 2.56 \\
 1000&    6.00 &  3.46 &  2.54  &  5.74 &  2.33 & 2.31 \\
 1200&    5.24 &  2.96 &  2.27  &  5.01 &  2.13 & 2.12 \\
 1500&    4.37 &  2.41 &  1.97  &  4.22 &  1.89 & 1.88 \\
 2000&    3.35 &  1.76 &  1.59  &  3.36 &  1.56 & 1.59 \\
\hline \hline
\end{tabular}
\end{center}
\caption{The total cross section and the total cross section for elastic
scattering and breakup reactions as obtained from the exact Faddeev
calculations are listed in the left part of the table. On the right side  the total
cross section as obtained from the Glauber calculation (including the second order
correction) is given. The total cross sections for breakup reactions,
$\sigma_{br}$, is
obtained by subtracting the total cross sections ((a) and (b)) for elastic
scattering taken from
Table~\ref{table2} from the total cross section $\sigma_{tot}^{\rm Gl}$.
} \label{table3}
\end{table}

\clearpage

\noindent

\end{document}